\documentclass[useAMS,twocolumn,usenatbib,usegraphicx]{mn2e}
\usepackage{graphicx}

\newcommand{\go}{\mathrel{\raise.3ex\hbox{$>$}\mkern-14mu
             \lower0.6ex\hbox{$\sim$}}}
\newcommand{\lo}{\mathrel{\raise.3ex\hbox{$<$}\mkern-14mu
             \lower0.6ex\hbox{$\sim$}}}

\newcommand{\vecB}{\bmath B}
\newcommand{\vechatr}{\hat{\bmath r}}
\newcommand{\vechatm}{\hat{\bmath m}}
\newcommand{\vecr}{\bmath r}

\newcommand{\vecm}{\bmath m}
\newcommand{\vecn}{\bmath n}
\title{On frequency dependence of pulsar linear polarization}

\author[P. F. Wang, C. Wang, and J. L. Han]
{P. F. Wang\thanks{E-mail: pfwang@nao.cas.cn}, C. Wang, and J. L. Han \\
  National Astronomical Observatories, Chinese Academy of
  Sciences.  A20 Datun Road, Chaoyang District, Beijing 100012, China \\
}

\pagerange{\pageref{firstpage}--\pageref{lastpage}}

\begin{document}

\maketitle

\label{firstpage}

\begin{abstract}
Frequency dependence of pulsar linear polarization is investigated by
simulations of emission and propagation processes. Linearly polarized
waves are generated through curvature radiation by relativistic
particles streaming along curved magnetic field lines, which have
ordinary mode (O-mode) and extra-ordinary mode (X-mode) components. As
emitted waves propagate outwards, two mode components are separated
due to refraction of the O mode, and their polarization states are
also modified. According to the radius to frequency mapping, low
frequency emission is generated from higher magnetosphere, where
significant rotation effect leads the X and O modes to be
separated. Hence, the low frequency radiation has a large fraction of
linear polarization. As the frequency increases, emission is generated
from lower heights, where the rotation effect becomes weaker and the
distribution regions of two modes are more overlapped. Hence, more
significant depolarization appears for emission at higher
frequencies. In addition, refraction effect of the O mode becomes
serious in very deep magnetosphere, which bends the O mode emission
towards outer parts of a pulsar beam and also causes the separation of
mode distribution regions and hence the fractional linear polarization
increasing with frequency. If emission of different frequencies is
generated from a region of the same height, serious O mode refraction
can result in the decrease of both profile width and fractional linear
polarization. The observed frequency dependence of linear polarization
for some pulsars can be naturally explained within the scope of our
scenario.
\end{abstract}


\section{INTRODUCTION}

Pulsars are generally believed to be rapidly rotating neutron stars.
Their polarized emission has been observed over a wide frequency range
from about 100 MHz to 32GHz over more than forty years
\citep[e.g.][]{man71, wml+93, xkj+96, gl98, wj08,hdv+09, tjb+13}. Very
rich polarization features have been uncovered for both linear and
circular polarizations, e.g., `S'-shaped position angle curves,
orthogonal polarization modes, single sign and sign reversals of
circular polarization, and their evolutions with frequency,
etc. Linear polarization of average pulse profiles is predominant for
most pulsars at low frequencies. The polarization percentage generally
decreases with the increase of observing frequency
\citep{msc70,man71}. \citet{mth73} noticed that some pulsars exhibit a
critical frequency, below which the fractional linear polarization is
constant, but above which it decreases with increasing frequency. To
investigate the change of pulsar polarizations across a much wider
frequency range, \citet{xkj+96} extended polarization observations to
mm-wavelengths, and noticed significant depolarization at high
frequencies. Recently, \citet{jkm+08} pointed out that simple profiles
are more likely to be characterized by high linear polarization than
complex profiles, and polarization properties for different profile
components of a given pulsar behave differently with frequency, e.g.,
PSR J0922+0638.

In addition to analysis of polarization features for individual
pulsars, statistical investigations were also undertaken for pulsar
linear polarization. It has been noticed at early days that short
period pulsars tend to have high linear polarization
\citep{hmt71,mgs81}. Long period pulsars exhibit faster depolarization
over frequency than the short period ones
\citep{mgs81}. \citet{xsg+95} presented the anti-correlation between
the depolarization index and the acceleration potential near neutron
star surface. Several authors also demonstrated that the degree of
linear polarization is higher for a pulsar with a large spin-down
luminosity $\dot{E}$ \citep{qml+95,vkk98,cmk01,wj08}.

To understand various polarization features, numerous theoretical
researches on pulsar polarization have been developed on the emission
processes \citep[e.g.][]{xlh+00,drd10,wwh12,kg12} or the propagation
effects \citep[e.g.][]{wlh10,bp12}, both of which can lead to
depolarization. The depolarization of pulsar linear polarization is
generally attributed to simultaneous interaction of two modes of
orthogonally polarized radiation \citep{ms98,kjm05} or owing to
randomization of position angles in weaker magnetic fields in outer
magnetosphere \citep{mth75,mgs81}. For example, \citet{mc97}
attributed the depolarization to birefringence of the X mode and O
mode above pulsar polar caps. \citet{vlk98} qualitatively explained
depolarization at high frequencies by considering the evolution of one
natural plasma mode within pulsar magnetosphere. However, the
formation and distribution of the two modes within a pulsar beam were
not mentioned; the refraction effect was simply discussed by analogy
with the calcite crystal; the polarization limiting effect was simply
analysed; and the evolution of linear polarization across field line
planes was not investigated. To get a better understanding of the
frequency dependence of pulsar linear polarization, emission processes
together with propagation effects need to be investigated
systematically.

Recently, we jointly studied the polarized curvature radiation
together with propagation processes within pulsar magnetosphere
numerically \citep{wwh14}. We succeeded in demonstrating the
distributions of the X-mode and O-mode within a pulsar magnetosphere
and explained the depolarization across an entire pulsar beam. We
found that the depolarization is serious if the corotation of
relativistic particles is not considered. When the rotation is taken
into account, significant linear polarization will be
produced. However, the frequency dependence behaviour of linear
polarization was not incorporated in the analysis.

Stimulated by the observational and theoretical investigations, we
hereby develop the joint research of emission and propagation
processes to understand frequency dependence of pulsar linear
polarization by considering the polarized curvature radiation process
together with propagation effects in a pulsar magnetosphere. In
Section 2, we present theoretical basics for our
calculations. Linearly polarized radiation within an entire pulsar
beam, polarized pulse profile and their evolution with frequency are
described in Section 3. Comparisons with observations are elaborated
in Section 4. Discussions and conclusions are presented in Section 5.

%%%%%%%%%%%%%%%%%%%%%%%%%%%%%%%%%%%%%%%%%%%%%%%%%%%%%%%%%%%%%%%%%%%%%%%%%%
%%%%%%%%%%%%%%%%%%%%%%%%%%%%%%%%%%%%%%%%%%%%%%%%%%%%%%%%%%%%%%%%%%%%%%%%%%
%%%%%%%%%%%%%%%%%%%%%%%%%%%%%%%%%%%%%%%%%%%%%%%%%%%%%%%%%%%%%%%%%%%%%%%%%%
\section{Theoretical basics}

%%%%%%%%%%%%%%%%%%%%%%%%%%%%%%%%%%%%%%%%%%%%%%%%%%%%%%%%%%%%%%%%%%%%%%%%%%
%%%%%%%%%%%%%%%%%%%%%%%%%%%%%%%%%%%%%%%%%%%%%%%%%%%%%%%%%%%%%%%%%%%%%%%%%%
\subsection{Geometry and emission}

\begin{figure}
\centering
\includegraphics[angle=0, width=0.39\textwidth] {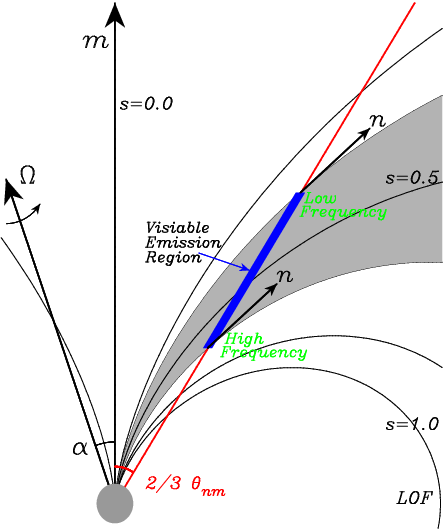}
\caption{Emission regions on a bundle of magnetic field lines with
  $\theta\simeq2/3\theta_{\vecn\vecm}$ for a given sight line. The
  magnetic moment $\vecm$ is inclined by an angle of $\alpha$ with
  respect to the rotation axis $\Omega$. The thick line outlines all
  tangential emission points of field lines.  $\theta_{\vecn\vecm}$
  represents the angle between the wave vector $\vecn$ and the
  magnetic momentum $\vecm$. The grey area represents the region for
  outflowing relativistic particles along a bundle of magnetic field
  lines. $s=0.0$ denotes the magnetic axis, $s=1.0$ denotes the last
  open field line (LOF).}
\label{fig:Emi_region}
\end{figure}

Pulsar radio emission is generally believed to be generated by
relativistic particles streaming out along the curved magnetic field
lines within the open magnetosphere. Due to the bending of field
lines, relativistic particles will experience perpendicular
acceleration and produce curvature radiation. Coherent curvature
radiation from bunches of relativistic particles serves as one of the
most probable mechanisms for pulsar radio emission. The coherent
particle bunch can be treated as a point-like huge charge for the sake
of simplicity. Pulsar magnetic fields are assumed to be of the static
dipole form in the radio emission region,
\begin{equation}
\vecB=B_{\star}(\frac{R_{\star}}{r})^3[3\vechatr(\vechatr\cdot
\vechatm)-\vechatm],
\label{eq:staticb}
\end{equation}
here $B_{\star}$ is magnetic field strength at the magnetic equator of
a neutron star, $R_{\star}$ is the neutron star radius, $\vechatr$ is
the unit vector along $\vecr$, and $\vechatm$ represents the unit
vector of the magnetic dipole moment. Within the static dipole
magnetosphere, emission generated from the regions of the same polar
angle $\theta$ within a given magnetic field line plane will point
towards the same direction, as shown in Fig.~\ref{fig:Emi_region}.

\begin{figure}
  \setlength{\tabcolsep}{0.5mm} \centering
  \begin{tabular}{cc}
    \includegraphics[angle=0, width=0.245\textwidth] {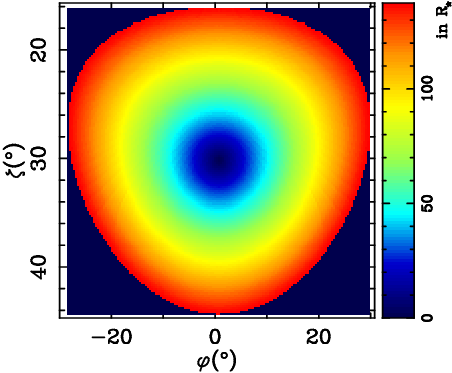} &
    \includegraphics[angle=0, width=0.245\textwidth] {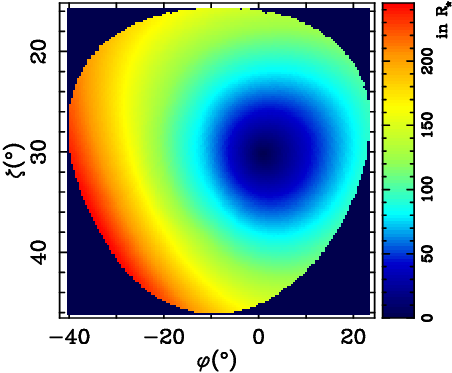} \\
    $\nu=0.8$GHz without rotation & $\nu=0.8$GHz with rotation\\
\\
    \includegraphics[angle=0, width=0.245\textwidth] {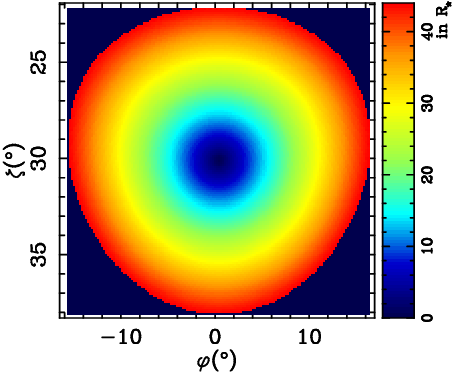} &
    \includegraphics[angle=0, width=0.245\textwidth] {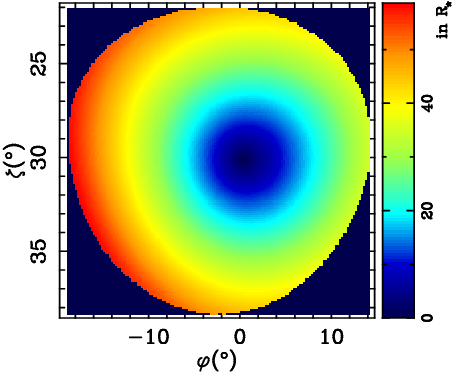} \\
    $\nu=1.4$GHz without rotation & $\nu=1.4$GHz with rotation\\
\\
    \includegraphics[angle=0, width=0.245\textwidth] {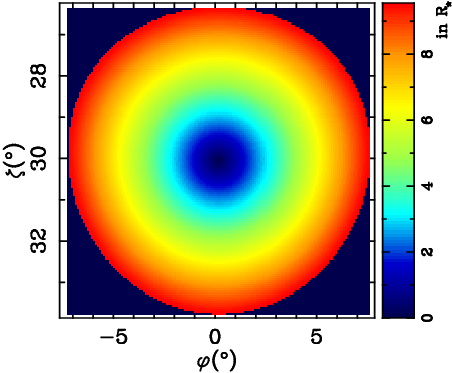} &
    \includegraphics[angle=0, width=0.245\textwidth] {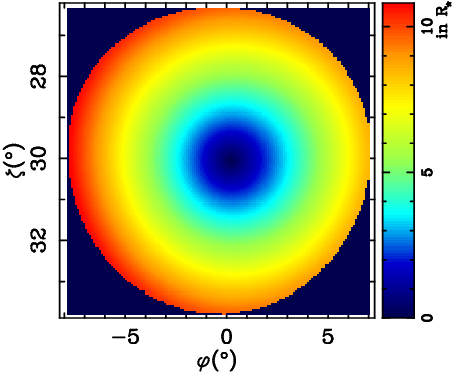} \\
    $\nu=3.0$GHz without rotation & $\nu=3.0$GHz with rotation\\
  \end{tabular}
  \caption{Emission heights (in the unit of $R_{\star}$) for different
    frequencies in a pulsar beam, plotted in the rotation phase
    $\varphi$ and the sight line angle $\zeta$. Corotation of
    relativistic particles is not considered for the left panels, but
    considered for the right panels. The parameters used for model
    calculations are an inclination angle of $\alpha=30^o$, a Lorentz
    factor of $\gamma=500$, and a rotation period of a pulsar $P=1\rm
    s$.}
  \label{fig:EHeights}
\end{figure}

Along a field line, emission radiated at different heights has
different frequencies. The characteristic frequency for curvature
radiation reads,
\begin{equation}
\nu_{\rm CR}=\frac{3 \gamma^3 c}{4\pi \rho}.
\label{eq:nucr}
\end{equation}
Here, $\gamma$ is the Lorentz factor of a relativistic particle,
$\rho$ represents the curvature radius for instaneous particle
trajectory. The larger curvature radius $\rho$ for particle trajectory
in the high magnetosphere naturally leads to lower frequency
emission. This is the physical basis for the radius to frequency
mapping.

For an inclined dipole magnetosphere, the pulsar emission beam is
generally compressed in the meridional plane defined by the rotation
axis and magnetic axis \citep{big90}. For simplicity, our calculations
assume that pulsar beams are circular, i.e., the last open field lines
which define the boundary of a pulsar beam have the same field line
constant of $r_{\rm e,lof}$, as defined by equation (15) of
\citet{gan04}. In general, the beam size in the meridional direction
does not affect our conclusions in this paper.

The emission regions at three different frequencies of $\nu_{\rm CR}$
are calculated as shown in Fig.~\ref{fig:EHeights}. The height ranges
are quasi symmetric about the magnetic axis for emission at a given
frequency generated from the static dipole magnetosphere.  Emission
near the beam centre comes from a lower altitude, as shown in the left
panels of Fig.~\ref{fig:EHeights}, because the curvature radii of the
inner field lines at a lower altitude are comparable to those of the
outer field lines at a higher altitude. Further, the entire emission
regions get deeper in the magnetosphere for a higher frequency, which
is caused by the smaller curvature radii of field lines in the deeper
magnetosphere. It should be noted that relativistic particles within a
pulsar magnetosphere not only stream along the curved magnetic field
lines, but also corotate with magnetosphere. When the corotation is
considered, the trajectories and velocities of relativistic particles
will be bent towards the rotation direction compared with those for
the static dipole magnetosphere. Hence, the emission regions become
asymmetry, with heights for the leading part of the beam getting
larger than those for the trailing part, as shown in the right panels
of Fig.~\ref{fig:EHeights}.

For emission of a single particle bunch at a given height in the
pulsar magnetosphere, the radiation field $\bmath E (t)$ and the
corresponding Fourier components $\bmath E (\omega)$ can be calculated
by using the circular path approximation, as described by
\citet{wwh12}. The emission from a relativistic particle bunch is
beamed in a $1/\gamma$ cone around the velocity direction. Therefore,
the detectable emission from a given height and rotation phase is
contributed by relativistic particles not only at the tangential
emission point of a field line but also on nearby field lines within
the $1/\gamma$ cone. By integrating the emission within an entire
emission cone from all field lines, we will obtain the total polarized
radiations in the given direction. In general, curvature radiation is
highly linearly polarized, and the rotation has great influences on
polarized curvature radiation process \citep{wwh14}.

%%%%%%%%%%%%%%%%%%%%%%%%%%%%%%%%%%%%%%%%%%%%%%%%%%%%%%%%%%%%%%%%%%%%%%%%%%
%%%%%%%%%%%%%%%%%%%%%%%%%%%%%%%%%%%%%%%%%%%%%%%%%%%%%%%%%%%%%%%%%%%%%%%%%%
\subsection{Propagation}

Once polarized emission is generated through the curvature radiation
process, they will be coupled to local plasma modes to propagate out
in a pulsar magnetosphere, which is filled with relativistic plasmas
that can be simply assumed to be cold (with a single Lorentz factor
$\gamma$) and with a density of $N_{\rm p}=\eta N_{\rm GJ}$. Here,
$\eta$ is the multiplicity factor and $N_{\rm GJ}$ represents the
Goldreich-Julian density \citep{gj69}. Two eigen transverse modes, the
X-mode and O-mode, of the plasma have been carefully investigated by
\citet{wlh10} and \citet{bp12}. The X-mode waves have a refraction
index of $n_{\rm X}=1$ and propagate rectilinearly; while the O-mode
waves have a refraction index of $n_{\rm O}<1$ and suffer refraction
\citep{ba86}. The polarization states of the X-mode and O-mode change
during their propagation. In the inner magnetosphere near the emission
region, the polarization states of both modes evolve adiabatically,
i.e., with the polarization of the X-mode wave orthogonal to the local
$\bmath k-\bmath B$ plane and the polarization of the O-mode wave
within the $\bmath k-\bmath B$ plane \citep{cr79}. As the waves
propagate to the higher magnetosphere, the adiabatic condition is not
satisfied and wave-mode coupling happens, i.e., one mode leaks to the
other and vice versa. The altitude where the mode coupling happens is
named as being the polarization limiting radius
\citep{cr79,wlh10}. Above the polarization limiting radius, the
polarization states of both modes are frozen and will not be further
affected by the plasma, except for possible cyclotron absorption in
the outer magnetosphere.

\begin{figure*}
  \centering
  \includegraphics[angle=0, width=0.87\textwidth] {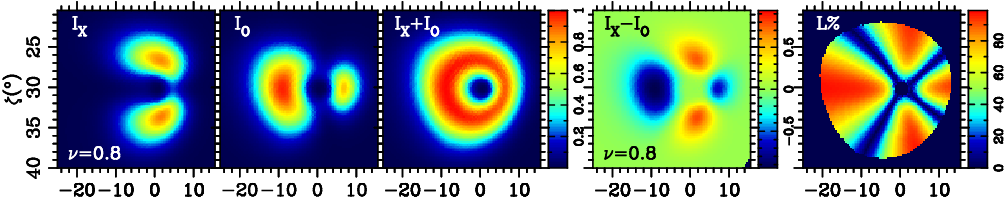} \\
  \includegraphics[angle=0, width=0.87\textwidth] {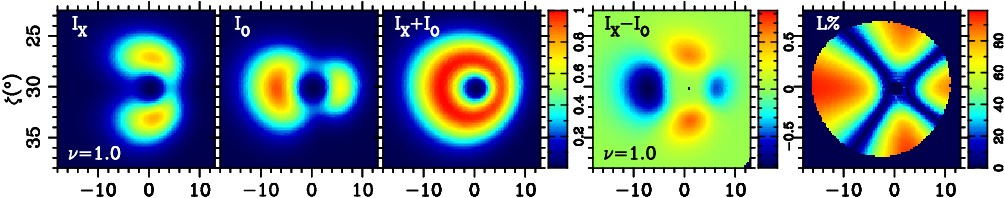} \\
  \includegraphics[angle=0, width=0.87\textwidth] {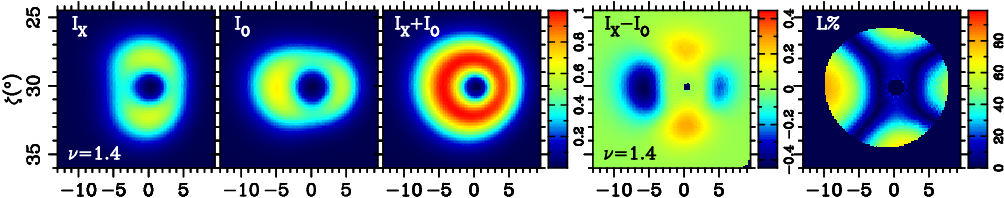} \\
  \includegraphics[angle=0, width=0.87\textwidth] {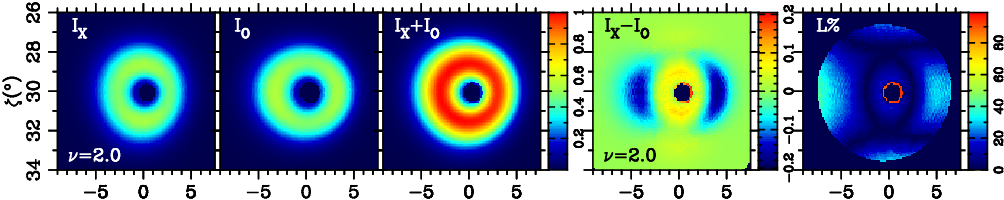} \\
  \includegraphics[angle=0, width=0.87\textwidth] {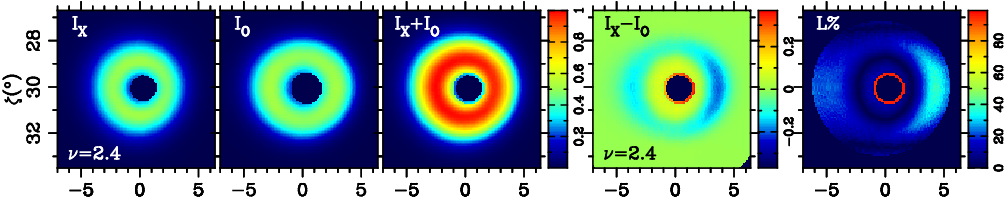} \\
  \includegraphics[angle=0, width=0.87\textwidth] {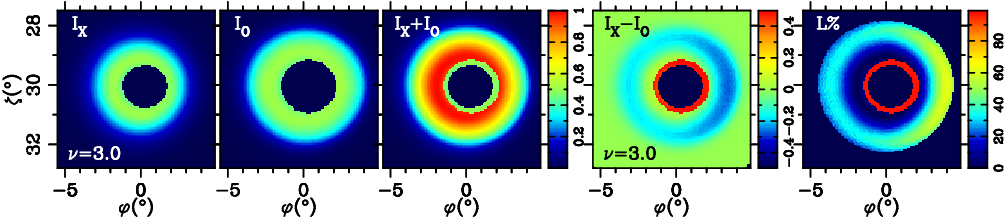} \\
  \caption{Pulsar polarization emission beams at frequencies of
    $\nu=$0.8GHz, 1.0GHz, 1.4GHz, 2.0GHz, 2.4GHz and 3.0GHz from top
    to bottom for the X-mode intensities ($I_{\rm X}$), the O-mode
    intensities ($I_{\rm O}$), the total intensity ($I_{\rm X}+I_{\rm
      O}$), their difference ($I_{\rm X}-I_{\rm O}$, i.e., the net
    linear polarization) and the fractional linear polarization
    ($L\%=|(I_{\rm X}-I_{\rm O})/(I_{\rm X}+I_{\rm
      O})|\times100\%$). The fractional linear polarization is plotted
    for the regions where the total intensity exceeds $1\%$ of the
    peak intensity. The density distribution for relativistic
    particles is assumed to be located in a cone at $s_p=0.5$ and with
    a width of $\sigma_s=0.12$. The other parameters used for model
    calculations are $\alpha=30^\circ$, $\gamma=500$, $\eta=100$,
    $B_{\star}=10^{12}\rm Gs$ and $P=1\rm s$.}
\label{fig:cone_beam}
\end{figure*}

\begin{figure*}
  \setlength{\tabcolsep}{0.5mm}
  \centering
  \begin{tabular}{cccc}
    \includegraphics[angle=0, width=0.25\textwidth] {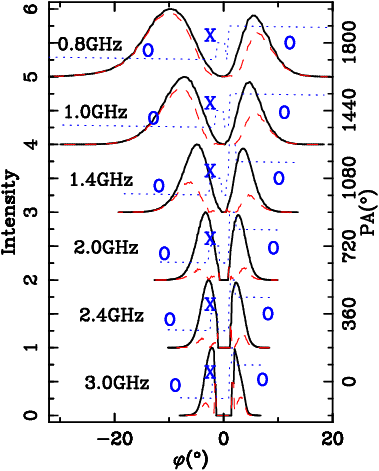} &
    \includegraphics[angle=0, width=0.25\textwidth] {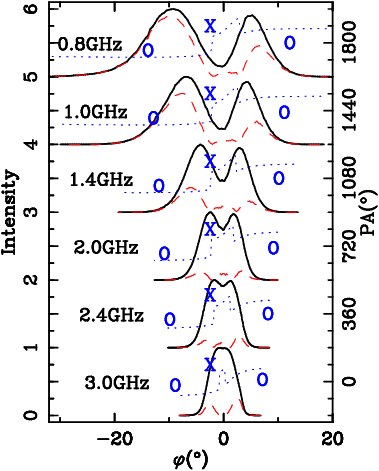} &
    \includegraphics[angle=0, width=0.25\textwidth] {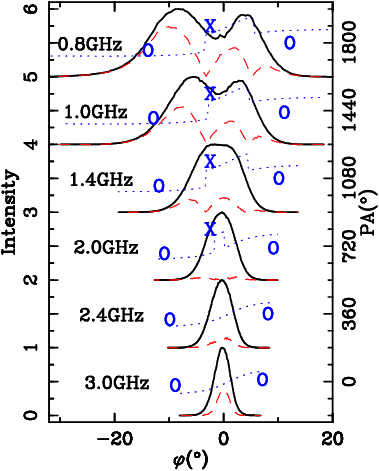} &
    \includegraphics[angle=0, width=0.25\textwidth] {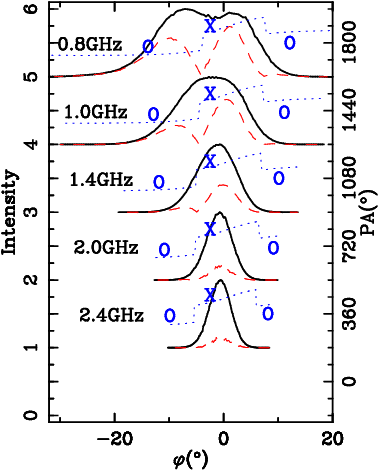} \\
    (a) $\zeta=30^o$ &  (b) $\zeta=31^o$ & (c) $\zeta=32^o$ & (d) $\zeta=33^o$\\
  \end{tabular}
  \caption{Polarized pulse profiles resulting from cutting of the
    beams in Fig.~\ref{fig:cone_beam} by the sight lines at
    $\zeta=30^o$, $31^o$, $32^o$ and $33^o$. The solid lines are for
    the total intensities, $I_{\rm X}+I_{\rm O}$, the dashed lines for
    the linear polarization intensities, $|I_{\rm X}-I_{\rm O}|$, and
    the dotted lines for position angles of linear polarization. The
    polarization modes (X or O) are marked near position angle
    curves.}
\label{fig:cone_profiles}
\end{figure*}

In addition to evolution of a single wave mode, the distributions of
polarized waves within each $1/\gamma$ emission cone also change
during propagation. As shown by \citet{wwh14}, the O mode refraction
will lead detectable X and O modes at a given position to be
incoherent, as they originate from different parts of a pulsar
beam. Adiabatic walking will result in the ordering of polarized waves
within the $1/\gamma$ cone, i.e., with the O-mode wave vectors
following the polarization plane of initial emission, but wave vectors
for the X-mode rotate by $90^o$. Therefore, incoherent superposition
of the orthogonal X-mode and O-mode emission causes depolarization for
total radiation power. If the corotation of relativistic particles is
not considered, significant depolarization occurs. When the corotation
is considered, the net polarization will reserve polarization features
of the stronger mode.

%%%%%%%%%%%%%%%%%%%%%%%%%%%%%%%%%%%%%%%%%%%%%%%%%%%%%%%%%%%%%%%%%%%%%%%%%%
%%%%%%%%%%%%%%%%%%%%%%%%%%%%%%%%%%%%%%%%%%%%%%%%%%%%%%%%%%%%%%%%%%%%%%%%%%
%%%%%%%%%%%%%%%%%%%%%%%%%%%%%%%%%%%%%%%%%%%%%%%%%%%%%%%%%%%%%%%%%%%%%%%%%%
\section{Emission at various frequencies}

After knowing emission regions and propagation effects, we can now
calculate the polarization states of emission at various
frequencies. The wave mode distributions within an entire pulsar beam
(the X mode intensity, $I_{\rm X}$, the O mode intensity, $I_{\rm O}$,
total intensity, $I_{\rm X}+I_{\rm O}$, and the intensity for the
linear polarization, $I_{\rm X}-I_{\rm O}$), polarized pulse profiles,
and their evolution with frequency can be calculated for the emission
generated from different heights for the entire open magnetosphere.

%%%%%%%%%%%%%%%%%%%%%%%%%%%%%%%%%%%%%%%%%%%%%%%%%%%%%%%%%%%%%%%%%%%%%%%%%%%
%%%%%%%%%%%%%%%%%%%%%%%%%%%%%%%%%%%%%%%%%%%%%%%%%%%%%%%%%%%%%%%%%%%%%%%%%%%
\subsection{Emission generated by particles with a single value of the Lorentz factor $\gamma$ at $\nu_{\rm CR}$}

In our calculations, the coherent particle bunch is simply treated as
a point-like huge charge. The emission intensity is closely related to
local particle densities. Here, the conal-shaped density distribution
for particles is adopted as $N(r,\theta,\phi)=N_0 r^{-3} f(\theta)
g(\phi)$, with
\begin{eqnarray}
f(\theta)&=&f_0\exp[-\frac{(s-s_p)^2}{2\sigma_s^2}],\nonumber\\
g(\phi)&=&1.
\label{eq:conal}
\end{eqnarray}
Here, $s=\theta_c/\theta_{c,max}$, $\theta_c$ is the polar angle of a
field line footed on the neutron star surface,
$\theta_{c,max}=\sin^{-1}(\sqrt{R_\star/r_{\rm e,lof}})$ represents
the polar angle maximum, i.e., the polar angle for the last open field
line footed on the neutron star surface, $s_p$ is the position for
density peak\footnote{In this paper the meaning of `s' is the same as
  $\vartheta$ in \citet{wwh14}.}.

Using such a particle density distribution, we can now get the
polarized emission at various frequencies in the model. The steps
follow \citet{wwh12,wwh14}. First, we find the field lines for
emission within the $1/\gamma$ cone pointing towards a given
direction. Then, the electric fields of the emission within the cone
will be decomposed into the X-mode and O-mode components following the
direction of local magnetic field. Evolution of the X-mode and O-mode
components within the $1/\gamma$ cone will be traced until the
polarization limiting radii by considering the refraction and the
adiabatic walking effects. The integration of emission within each
cone will result in the X-mode and O-mode intensities. Finally,
emission pointing outwards from the entire open field line region will
be mapped. The equations governing the ray trajectory and polarization
state evolutions have been given in details in \citet{wwh14}.

In Fig.~\ref{fig:cone_beam}, we show pulsar polarization emission
beams for the X-mode, O-mode, total intensity, and the linear
polarization at six frequencies of $\nu=$0.8GHz, 1.0GHz, 1.4GHz,
2.0GHz, 2.4GHz and 3.0GHz. The beams for low frequency emission are
generally larger than those for high frequencies due to a larger
opening angle from a higher altitude. Moreover, the X-mode and O-mode
dominate different parts of the pulsar beam, i.e., the X-mode
dominates at two sides of the beam in $\zeta$ direction while the
O-mode dominates in $\phi$ direction, as shown in the panels for $I_X$
and $I_O$ in Fig.~\ref{fig:cone_beam}. Further, emission regions for
the X mode and O mode are very distinct for lower frequency emission
coming from a higher altitude. With increase of observing frequency,
the distribution regions of the two modes gradually overlap, and hence
serious depolarization happens, as shown by $I_X-I_O$ and fractional
linear polarization of $L\%$ in Fig.~\ref{fig:cone_beam}. The
differences are caused mainly by different corotation velocities of
relativistic particles. When the corotation of relativistic particles
is not considered, the distribution regions for two modes are almost
completely overlapped \citep{wwh14}. The O-mode distribution regions
are generally broader than those for the X-mode for the beams at
higher frequencies, as shown in Fig.~\ref{fig:cone_beam}, which is
caused by the dense plasma in deeper magnetosphere where serious
refraction effect results in a prominent outward shift of the O mode
beam.

\begin{figure}
  \centering
  \includegraphics[angle=0, width=0.28\textwidth] {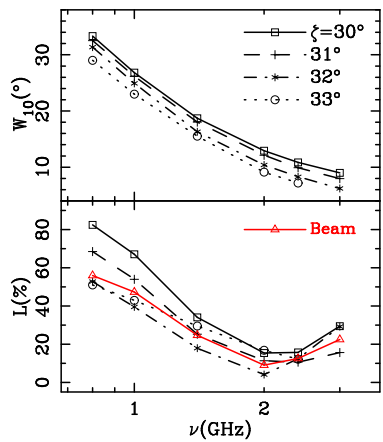}
  \caption{Evolution of fractional linear polarization $L\%$ and
    profile width $W_{10}$ with frequency. The fractional linear
    polarizations are calculated for the profiles in
    Fig.~\ref{fig:cone_profiles} with the sight lines of $\zeta=30^o$,
    $31^o$, $32^o$ and $33^o$ at frequencies $\nu=0.8$, $1.0$, $1.4$,
    $2.0$, $2.4$ and $3.0$ GHz. The red solid line represent the
    fractional linear polarizations for the entire beams in
    Fig.~\ref{fig:cone_beam} with the total intensity above $1\%$ the
    peak intensity. The model parameters are the same as those in
    Fig.~\ref{fig:cone_beam}, i.e., $\alpha=30^\circ$, $\gamma=500$,
    $\eta=100$, $B_{\star}=10^{12}\rm Gs$ and $P=1\rm s$.}
  \label{fig:LPfrac_nu}
\end{figure}

When a given sight line cuts across a pulsar beam, it will find
polarized pulse profiles at a series of frequencies, as shown in
Fig.~\ref{fig:cone_profiles}. For the central cut at $\zeta=30^o$, the
observed pulse profiles are predominantly of the O mode except for a
small central phase region for the X mode. Emission at a lower
frequency, like $\nu=0.8$GHz, is highly linearly polarized. With the
increase of frequency, the fractional linear polarization, $L\%$,
gradually decreases and the profile width, $W_{10}$, narrows down
correspondingly, as shown in Fig.~\ref{fig:LPfrac_nu}. As observing
frequency further increases to about 3.0GHz, the pulse profile keeps
narrowing down, but the fractional linear polarization increases. This
is because the refraction effect becomes stronger as the plasma
density increases, which dominates over the influence of rotation in
determining pulsar linear polarization. The frequency dependencies of
fractional linear polarization and profile width are similar for
different sight lines, such as $\zeta=31^o$, $32^o$ and
$\zeta=33^o$. Moreover, the fractional linear polarization of the
entire pulse beam exhibits a similar frequency dependence as those for
the above discussed pulse profiles, which is outlined by the red solid
line in Fig.~\ref{fig:LPfrac_nu}.

\begin{figure}
  \setlength{\tabcolsep}{0.5mm}
  \centering
  \begin{tabular}{cc}
    \includegraphics[angle=0, width=0.235\textwidth] {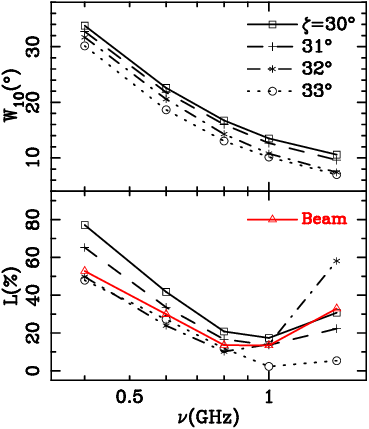} &
    \includegraphics[angle=0, width=0.235\textwidth] {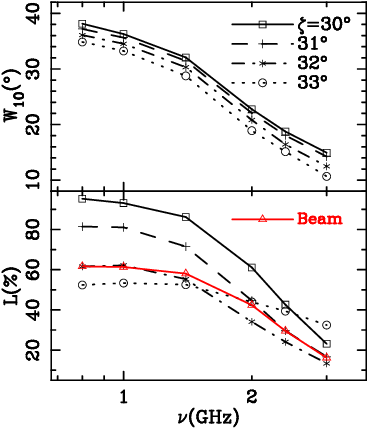} \\
    (a) $\gamma=400$, $\eta=100$& (b) $\gamma=600$, $\eta=100$ \\
    \\
    \includegraphics[angle=0, width=0.235\textwidth] {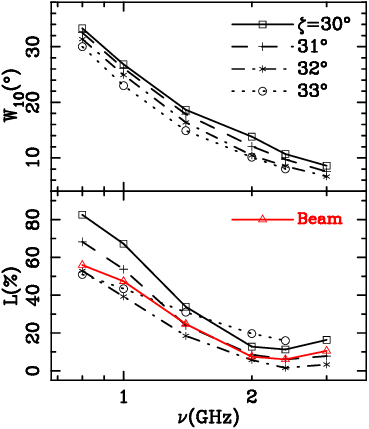} &
    \includegraphics[angle=0, width=0.235\textwidth] {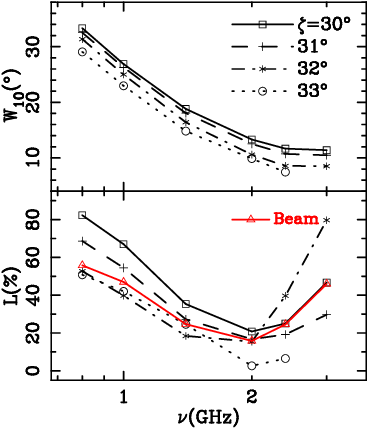} \\
    (c) $\gamma=500$, $\eta=10$ & (d) $\gamma=500$, $\eta=1000$ \\
  \end{tabular}
  \caption{Same as Fig.~\ref{fig:LPfrac_nu} except for particles with
    different Lorentz factor $\gamma$ or density $\eta$.}
\label{fig:LPfrac_nu_diff}
\end{figure}

The frequency dependencies of fractional linear polarization and
profile width are also calculated for plasmas with various Lorentz
factors $\gamma$ and multiplicity parameters $\eta$, as shown in
Fig.~\ref{fig:LPfrac_nu_diff}. It can be seen from
Fig.~\ref{fig:LPfrac_nu_diff}(a) that profile widths decrease with
the increase of frequency, but the fractional linear polarization
decreases first and then may increase, similar to those in
Fig.~\ref{fig:LPfrac_nu}, except that the lower frequency emission
is generated by relativistic particles with a smaller Lorentz
factor. If particles have larger Lorentz factors, emission of a
given frequency will be generated in higher magnetosphere. Because
the rotation has already separated the distribution regions for the
X and O modes above a certain height, emission above the height will
be highly polarized with an almost constant polarization degree, as
shown in Fig.~\ref{fig:LPfrac_nu_diff}(b). We can know from
Fig.~\ref{fig:LPfrac_nu_diff}(c) and (d) that high frequency
emission generated in a dense plasma with a large $\eta$ in deeper
magnetosphere tends to have a larger fraction of linear polarization
than those in less dense plasma with a small $\eta$. This is caused
by mode separation induced by the O-mode refraction, the effect is
proportional to $\eta$ \citep{ba86}.

In summary, with curvature radiation mechanism, the radiation keeps
highly linearly polarized at low frequencies. The polarization
gradually decreases with observing frequency, but may increase at a
very high frequency. The rotation plays an important role in keeping
high degree of linear polarization, and can cause the decrease of
fractional linear polarization in higher magnetosphere. Serious
refraction in a deep magnetosphere, which is significant when the
plasma is dense and particles have a small Lorentz factor, will lead a
larger fraction of linear polarization for high frequency emission.

%%%%%%%%%%%%%%%%%%%%%%%%%%%%%%%%%%%%%%%%%%%%%%%%%%%%%%%%%%%%%%%%%%%%%%%%%%
%%%%%%%%%%%%%%%%%%%%%%%%%%%%%%%%%%%%%%%%%%%%%%%%%%%%%%%%%%%%%%%%%%%%%%%%%%
\subsection{Emission generated from a region of the same height}

\begin{figure*}
  \centering
  \includegraphics[angle=0, width=0.87\textwidth]{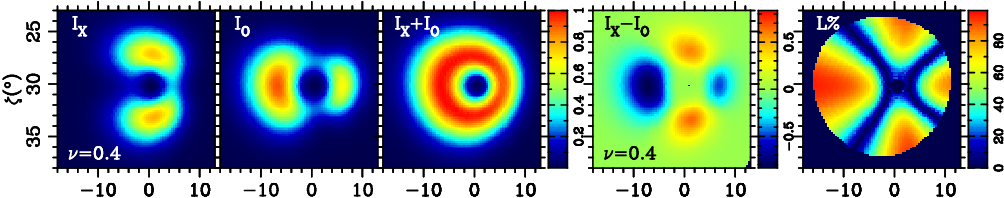} \\
  \includegraphics[angle=0, width=0.87\textwidth]{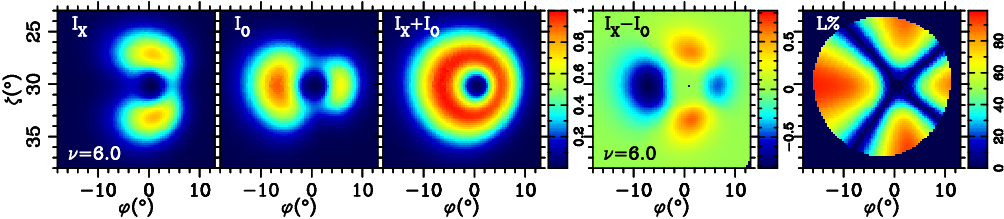} \\
  (a) $\gamma=500$, $\eta=100$\\
  \includegraphics[angle=0, width=0.87\textwidth]{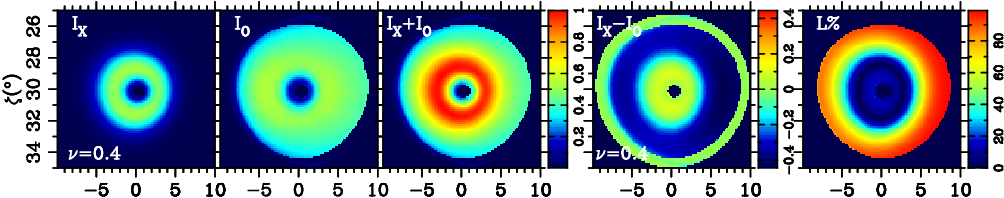} \\
  \includegraphics[angle=0, width=0.87\textwidth]{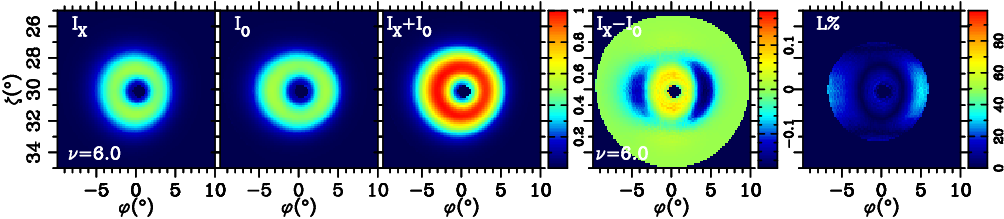} \\
  (b) $\gamma=400$, $\eta=1000$\\
  \caption{Same as Fig.~\ref{fig:cone_beam} for polarization beams but
    with different parameters. Panels (a) are for the plasma with
    $\gamma=500$ and $\eta=100$. Emission at frequencies of 0.4 and
    6.0GHz is generated from the same height region as the emission of
    $\nu_{\rm CR}=1.0$ GHz. Panels (b) are for the plasma with
    $\gamma=400$ and $\eta=1000$. The other parameters used for model
    calculations are the same as those in Fig.~\ref{fig:cone_beam},
    i.e., $\alpha=30^\circ$, $B_{\star}=10^{12}\rm Gs$ and $P=1\rm
    s$.}
\label{fig:cone_beam_one}
\end{figure*}

We have investigated the frequency dependence of pulsar linear
polarization emitted at the characteristic frequency $\nu_{\rm
  CR}$. Curvature radiation from a relativistic particle or a particle
bunch has a spectrum which is a power law with an index of $1/3$ at
the lowest frequency and decreases exponentially at high frequencies,
not just at the frequency of $\nu_{\rm CR}$. Hence, emission of
various frequencies may be produced from the region of the same height
in a pulsar magnetosphere.

\begin{figure}
  \setlength{\tabcolsep}{0.5mm}
  \centering
  \begin{tabular}{cc}
    \includegraphics[angle=0, width=0.239\textwidth] {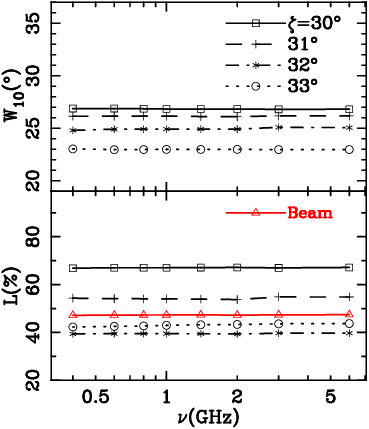} &
    \includegraphics[angle=0, width=0.239\textwidth] {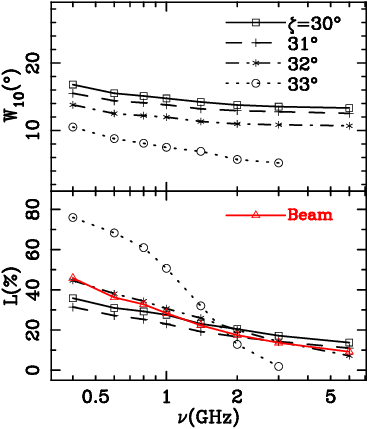} \\
    (a) $\gamma=500$, $\eta=100$ &  (b) $\gamma=400$, $\eta=1000$  \\
  \end{tabular}
  \caption{Same as Fig.~\ref{fig:LPfrac_nu} but for (a) Lorentz factor
    $\gamma=500$ and plasma density $\eta=100$, and (b) Lorentz factor
    $\gamma=400$ and plasma density $\eta=1000$. The fractional linear
    polarization and profile width are calculated by cutting the beams
    in Fig.~\ref{fig:cone_beam_one}.}
\label{fig:LPfrac_nu_one}
\end{figure}

For investigations conducted here, it is assumed that emission of
different frequencies is generated from the region of the same height
as the emission of $\nu_{\rm CR}=1.0$GHz. For a given $\nu_{\rm CR}$,
the shape of curvature radiation spectrum can be uniquely
determined. Therefore, emission of different frequencies has the same
beam pattern. However, due to the influence of frequency dependent
propagation effects, the final beam patterns will be modified. As done
in the previous section, we calculate intensity distributions of the
two mode within the entire pulsar beam, and then polarized pulse
profiles for various frequencies and sight line angles, and finally
find the corresponding frequency dependencies of fractional linear
polarization and profile width, as shown in
Figs.~\ref{fig:cone_beam_one} and \ref{fig:LPfrac_nu_one}.

For moderate plasma conditions with a Lorentz factor of $\gamma=500$
and a density of $\eta=100$, the wave mode distributions within a
pulsar beam are similar for various frequencies, as shown in
Fig.~\ref{fig:cone_beam_one} (a). The profile widths and fractional
linear polarizations almost do not change with frequency, as shown in
Fig.~\ref{fig:LPfrac_nu_one} (a). This kind of frequency dependence is
caused by the fact that the effects of rotation are the same for
emission of different frequencies, because they are generated from the
same height region, and the O mode components are less affected by the
refraction effect. If the plasma has a smaller Lorentz factor of
$\gamma=400$ but with a larger density of $\eta=1000$, the
distribution regions for the X and O modes are overlapped, and the
refraction effect becomes much more significant, especially for lower
frequency emission \citep{ba86,wwh14}. It results in broader
distribution regions for O mode than those for X mode, as shown in
Fig.~\ref{fig:cone_beam_one} (b). Compared with emission at $6.0$GHz,
the O mode distribution region for emission at $0.4$ GHz is even much
broadened. Therefore, both the profile width and fractional linear
polarization decrease with frequency. This frequency dependence is due
to that polarization features at higher frequencies are generated in
deeper magnetosphere within a small altitude range. Our result is
contrary to the argument given by \citet{mc97} that the refraction
causes average pulse profiles to depolarize while the profile width
remains a constant.

%%%%%%%%%%%%%%%%%%%%%%%%%%%%%%%%%%%%%%%%%%%%%%%%%%%%%%%%%%%%%%%%%%%%%%%%%%
%%%%%%%%%%%%%%%%%%%%%%%%%%%%%%%%%%%%%%%%%%%%%%%%%%%%%%%%%%%%%%%%%%%%%%%%%%
\subsection{Emission generated by particles with $\gamma$ distribution}

\begin{figure}
  \setlength{\tabcolsep}{0.4mm}
  \centering
  \includegraphics[angle=0, width=0.28\textwidth] {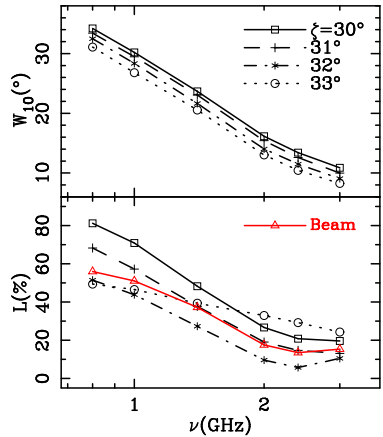}
  \caption{Same as Fig.~\ref{fig:cone_beam} except for particles with
    a gaussian distribution of energy, with $\gamma_m=500$, and
    $\sigma_{\gamma}=50$. }
  \label{fig:LPfrac_nu_add}
\end{figure}

The frequency dependence of pulsar emission has been considered for
curvature radiation of relativistic particles with a single $\gamma$
in the regions with either a range of heights or the same
height. Relativistic particles within a pulsar magnetosphere generated
by the sparking process may have an energy distribution \citep{ml10},
which we assume to be $N(\gamma) \sim \exp[-(\gamma-\gamma_m)/2
  \sigma_{\gamma}^2]$. Doing the same as previous subsections, we get
intensities and polarizations of wave modes within the entire pulsar
beam for each discrete Lorentz factor, and then sum them incoherently
for total radiation. The polarized pulse profiles and their frequency
dependencies are calculated, as shown in Fig.~\ref{fig:LPfrac_nu_add},
which are similar to those for the relativistic particles with a
single $\gamma$ shown in Fig.~\ref{fig:cone_beam}. The frequency
dependence here becomes less steep. Depolarization, which is caused by
addition of emission by relativistic particles with different Lorentz
factors, leads emission at 3.0GHz to be less polarized compared to
emission produced by particles with a single $\gamma$.

%%%%%%%%%%%%%%%%%%%%%%%%%%%%%%%%%%%%%%%%%%%%%%%%%%%%%%%%%%%%%%%%%%%%%%
%%%%%%%%%%%%%%%%%%%%%%%%%%%%%%%%%%%%%%%%%%%%%%%%%%%%%%%%%%%%%%%%%%%%%%
%%%%%%%%%%%%%%%%%%%%%%%%%%%%%%%%%%%%%%%%%%%%%%%%%%%%%%%%%%%%%%%%%%%%%%
\section{Comparison with observations}

We have numerically simulated wave mode distribution regions and
polarization beams, calculated polarized pulse profiles and their
evolution with frequency for the conal-shaped density model with
various plasma conditions. We have demonstrated the frequency
dependent depolarization without lose of generality. In general, the
fractional linear polarization together with profile width decrease
with frequency. In order to check the validity of our simulations, it
is worthwhile to compare our results with pulsar polarization
observations.

\begin{figure}
 \setlength{\tabcolsep}{0.5mm}
  \centering
  \begin{tabular}{cc}
  \includegraphics[angle=0, width=0.235\textwidth] {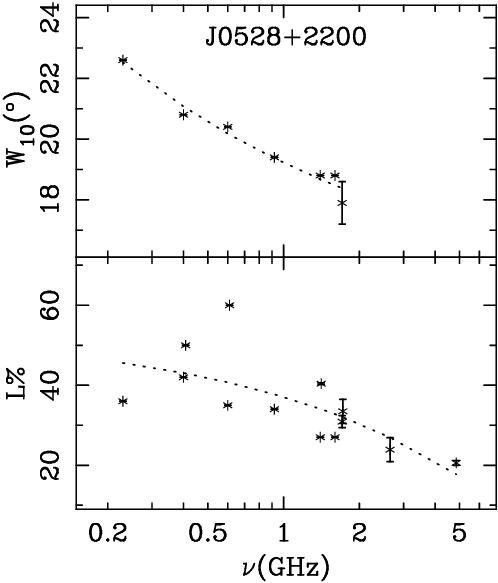} &
  \includegraphics[angle=0, width=0.235\textwidth] {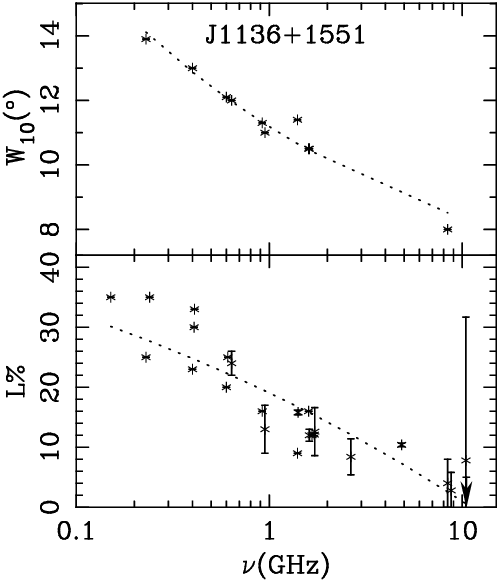} \\
  \smallskip
  \includegraphics[angle=0, width=0.235\textwidth] {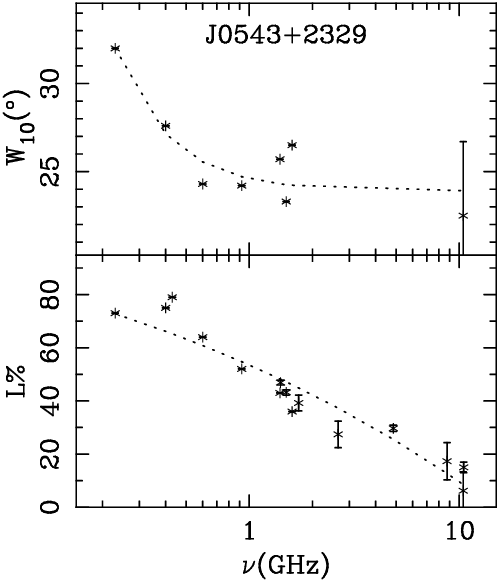} &
  \includegraphics[angle=0, width=0.235\textwidth] {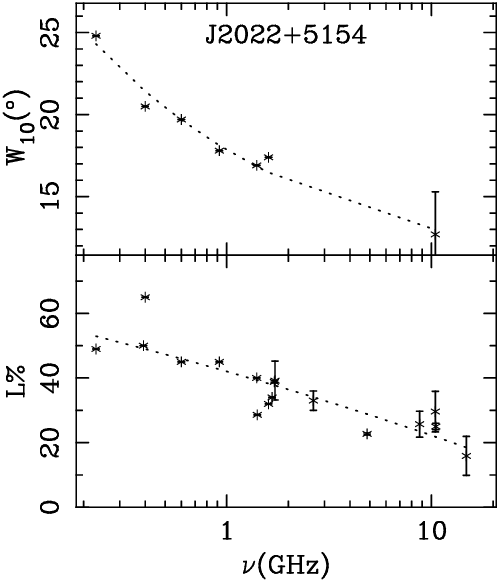} \\
  \smallskip
  \includegraphics[angle=0, width=0.235\textwidth] {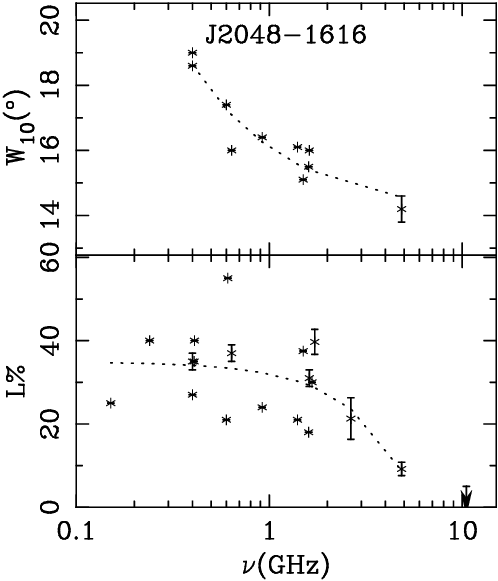} &
  \includegraphics[angle=0, width=0.235\textwidth] {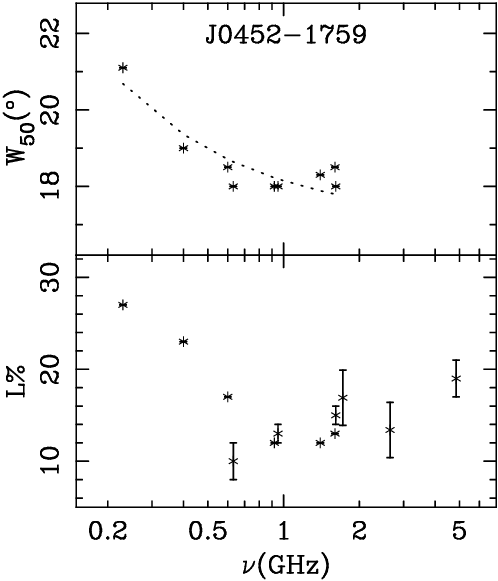} \\
  \end{tabular}
  \caption{The profile widths, $W_{10}$, and fractional linear
    polarizations, $L\%$, at a series of frequencies. The data are
    collected from literatures as listed in Table \ref{twl}. The dotted
    lines represent model fitting.}
\label{fig:psr_observ}
\end{figure}

Fig.~\ref{fig:psr_observ} shows the changes of observed profile widths
and fractional linear polarizations for six pulsars at a series of
frequencies, with the data collected in table~\ref{twl} in appendix
from literatures. These pulsars are of different morphological types
\citep{ran83,lm88}. {\bf PSR J0528+2200} and {\bf J1136+1551} are
conal double pulsars, each of which has two well separated conal
components and smoothly varied position angle curves. {\bf PSR
  J0543+2329} and {\bf J2022+5154} belong to the type of partial cone,
which has only one conal component and represents one side of conal
emission as indicated by position angle curve. {\bf PSR J2048-1616} is
a triple-component pulsar. It has two conal components at the outer
parts of pulse profiles and one core component in middle. With
increasing frequency, the central core component gradually disappears
as it has a steeper spectrum \citep{ran83}. {\bf PSR J0452-1759} has
multiple components.

The evolution of profile width and fractional linear polarization with
frequency can be interpreted for these pulsars by combing the
curvature radiation processes together with propagation effects within
pulsar magnetosphere. The widths of all these pulsars decrease with
frequency. A constant plus a power law, $W_{10}=W_c+(\nu/\nu_0)^{a}$,
can account for the frequency dependence, as pointed out by
\citet{tho91}. The frequency dependence of pulse width or beam width
can be understood in the frame of curvature radiation mechanism by
involving the density and energy distributions of relativistic
particles within a pulsar magnetosphere \citep{whw13}. The fractional
linear polarization is found to decreases with frequency as well. A
constant minus a power law, $L=L_c-(\nu/\nu_0)^{b}$, can be used to
fit the frequency dependence, as indicated by dotted lines in
Fig.~\ref{fig:psr_observ}, which could result from depolarization by
addition of the X mode and O mode as discussed in section 3.1 and
3.2. However, the fractional linear polarization of PSR J0452-1759
decreases first and then increases, and the increase of fractional
linear polarization may be due to the significant refraction effect,
as discussed in section 3.1.

%%%%%%%%%%%%%%%%%%%%%%%%%%%%%%%%%%%%%%%%%%%%%%%%%%%%%%%%%%%%%%%%%%%%%%
%%%%%%%%%%%%%%%%%%%%%%%%%%%%%%%%%%%%%%%%%%%%%%%%%%%%%%%%%%%%%%%%%%%%%%
%%%%%%%%%%%%%%%%%%%%%%%%%%%%%%%%%%%%%%%%%%%%%%%%%%%%%%%%%%%%%%%%%%%%%%
\section{Discussions and Conclusions}

In this paper, we have investigated the frequency dependence of pulsar
linear polarization by considering the curvature radiation process
together with propagation effects. The X and O mode distributions
within the entire pulsar beam, polarized pulse profiles and their
evolution with frequency are calculated for emission at different
frequencies originated from regions of different or the same height in
a pulsar magnetosphere. We find the following conclusions:

(i) Rotation can lead to the separation of distribution regions for
the X and O mode within a pulsar beam, the effect of which is much
more significant for low frequency emission generated at higher
magnetosphere.

(ii) The fractional linear polarization tends to approach a constant
degree towards low frequency limit, as the distribution regions of the
two modes have almost been separated above certain heights.

(iii) The fractional linear polarization generally decreases with
frequency. Because the higher frequency emission is generated from a
lower altitude, where rotation induced mode separation is not
significant.

(iv) The fractional linear polarization could increase at higher
frequency due to the significant O mode refraction.

(v) If emission originates from almost the same height region, the O
mode refraction will lead to the decrease of profile width and
fractional linear polarization with frequency. The refraction
influence is much more significant if the plasma has a large density
or a small Lorentz factor.

(vi) Depolarization caused by the randomization of polarized emission
from different heights of pulsar magnetosphere is obvious for high
frequency emission generated from the hot relativistic
plasma. However, it does not change the basic tendency of which the
fractional linear polarization decrease with frequency.

The model simulations in this paper may help us to understand some
statistical features of pulsar linear polarization. For short period
pulsars, relativistic particles within the magnetosphere may suffer
much more significant rotation influences. Hence, the X mode and O
mode distribution regions tend to be separated and the pulsars may
have higher linear polarization. The relativistic particles of high
$\dot{E}$ pulsars may be accelerated to a very higher energy (larger
$\gamma$). Their emission can be generated from a much higher
magnetosphere, where the significant rotation will also lead to a
large fraction of linear polarization.

In our calculations, the frequency dependence of emission region is
determined by referring to the characteristic frequency of curvature
radiation, and emission intensities are closely related to local
plasma density. However, emission intensities depend on the detailed
coherent process within pulsar magnetosphere, which is not yet
resolved now. Moreover, our calculations simply assume that streams of
relativistic plasmas are cold and distributed in a conal shaped
region, which travel along the curved dipole magnetic field lines and
corotate with a pulsar magnetosphere. However, the actual energy and
density distributions of particles are unknown, which can be hot and
with irregular density shapes \citep{ml10}. Furthermore, due to the
influences of rotation and polar cap current, pulsar magnetosphere may
be of distorted dipole form \citep{kg13}, which may affect pulsar
polarization states and is not yet considered in this paper.

Pulsars exhibit rich diversities of polarization profiles, some of
which are not consistent with our model predictions. For example, the
profile width of PSR J1239+2453 decreases with observing frequency,
but its fractional linear polarization increase first and then
decrease; the profile widths of PSR J0601-0527 decrease but their
fractional linear polarizations increase; the profile width and
fractional linear polarization for PSR J2113+4644 increase. Moreover,
individual profile components of a given pulsar can behave differently
with frequency \citep{jkm+08}. For example, the highly polarized
profile component of PSR J0922+0638 remain highly polarized at high
frequencies. These diverse polarization features need further
investigations.

%%%%%%%%%%%%%%%%%%%%%%%%%%%%%%%%%%%%%%%%%%%%%%%%%%%%%%%%%%%%%%%%%%%%
\section*{Acknowledgements}

This work has been supported by the National Natural Science
Foundation of China (11403043, 11473034 and 11273029), and the
Strategic Priority Research Program `The Emergence of Cosmological
Structures' of the Chinese Academy of Sciences (Grant
No. XDB09010200).

%%%%%%%%%%%%%%%%%%%%%%%%%%%%%%%%%%%%%%%%%%%%%%%%%%%%%%%%%%%%%%%%
%%%%%%%%%%%%%%%%%%%%%%%%%%%%%%%%%%%%%%%%%%%%%%%%%%%%%%%%%%%%%%%%
%%%%%%%%%%%%%%%%%%%%%%%%%%%%%%%%%%%%%%%%%%%%%%%%%%%%%%%%%%%%%%%%
\bibliographystyle{mn2e}

\section*{Appendix}
\begin{table}
  \scriptsize
  \begin{center}
    \caption{The profile width, $W_{10}$, and the fractional linear
      polarization, $L\%$, at various frequencies for six pulsars.}
    \label{twl}
    \scriptsize
    \begin{tabular}{cllll}
      \hline
      \hline
      PSR Jname  & Freq.(GHz) & $W_{10}(^o)$ & $L\%$ & Ref  \\
      \hline
J0452-1759$^\dagger$ &    0.23    & 21.1        &  27         & 1 \\
                 &    0.4     & 19          &  23         & 1 \\
                 &    0.6     & 18.5        &  17         & 1 \\
                 &    0.631   & 18          & $10\pm2$    & 6 \\
                 &    0.92    & 18          &  12         & 1  \\
                 &    0.95    & 18          & $13\pm1$    & 10 \\
                 &    1.4     & 18.3        &  12         & 1 \\
                 &    1.6     & 18.5        &  13         & 1 \\
                 &    1.612   & 18          & $15\pm1$    & 7 \\
                 &    1.72    & -           & $16.9\pm3$  & 8 \\
                 &    2.65    & -           & $13.4\pm3$  & 8 \\
                 &    4.85    & -           & $19\pm2$    & 11 \\
      \hline
      J0528+2200 &    0.23    & 22.6        &  36         & 1 \\
                 &    0.4     & 20.8        &  42         & 1 \\
                 &    0.408   &   -         &  50         & 4 \\
                 &    0.6     & 20.4        &  35         & 1 \\
                 &    0.61    &   -         &  60         & 4 \\
                 &    0.92    & 19.4        &  34         & 1 \\
                 &    1.4     & 18.8        &  27         & 1 \\
                 &    1.41    &   -         &  $40.4\pm0.2$ & 12 \\
                 &    1.6     & 18.8        &  $27$         & 1 \\
                 &    1.71    & $17.9\pm0.7$&  $30.9\pm1.5$ & 12 \\
                 &    1.72    &   -         &  $33.5\pm3$   & 8 \\
                 &    2.65    &   -         &  $23.9\pm3$   & 8 \\
                 &    4.85    &   -         &  $20.6\pm0.6$ & 12 \\
      \hline
      J0543+2329 &    0.23    &   32.0      &  73         & 1 \\
                 &    0.4     &   27.6      &  75         & 1 \\
                 &    0.43    &     -       &  79         & 9 \\
                 &    0.6     &   24.3      &  64         & 1 \\
                 &    0.92    &   24.2      &  52         & 1 \\
                 &    1.4     &   25.7      &  43         & 1 \\
                 &    1.41    &     -       &  $46.9\pm0.9$ & 12 \\
                 &    1.5     &   23.3      &  $43.2\pm1.0$ & 13 \\
                 &    1.6     &   26.5      &  $36$         & 1 \\
                 &    1.72    &     -       &  $39.2\pm3$   & 8 \\
                 &    2.65    &     -       &  $27.4\pm5$   & 8 \\
                 &    4.85    &     -       &  $29.7\pm1.2$ & 12 \\
                 &    8.75    &     -       &  $17.3\pm7$   & 8 \\
                 &    10.45   & $22.5\pm4.2$&  $6.2\pm7.1$  & 12 \\
                 &    10.55   &     -       &  $15\pm2$     & 14 \\
      \hline
      J1136+1551 &    0.151   &    -        &  35         & 4 \\
                 &    0.23    &   13.9      &  25         & 1 \\
                 &    0.24    &    -        &  35         & 4 \\
                 &    0.4     &   13.0      &  23         & 1 \\
                 &    0.408   &    -        &  30         & 4 \\
                 &    0.41    &    -        &  33         & 5 \\
                 &    0.6     &   12.1      &  20         & 1 \\
                 &    0.61    &    -        &  25         & 4 \\
                 &    0.638   &   12        &  $24\pm2$   & 6 \\
                 &    0.92    &   11.3      &  16         & 1 \\
                 &    0.95    &   11        &  $13\pm4$   & 10 \\
                 &    1.4     &   11.4      &  9          & 1 \\
                 &    1.41    &     -       &  $15.8\pm0.4$ & 12 \\
                 &    1.6     &   10.5      &  $16$         & 1 \\
                 &    1.612   &   10.5      &  $12\pm1$     & 7 \\
                 &    1.665   &    -        &  $12$         & 5 \\
                 &    1.71    &    -        &  $12.1\pm0.2$ & 12 \\
                 &    1.72    &     -       &  $12.6\pm4$   & 8 \\
                 &    2.65    &     -       &  $8.4\pm3$    & 8 \\
                 &    4.85    &     -       &  $10.4\pm0.3$ & 12 \\
                 &    8.4     &     8       &  $4\pm4$      & 3 \\
                 &    8.75    &     -       &  $2.8\pm3$    & 8 \\
                 &    10.45   &     -       &  $7.8\pm23.9$ & 12 \\
                 &    10.55   &     -       &  $\leq5$      & 14 \\
      \hline
      J2048-1616 &    0.151   &    -        &  25         & 4 \\
                 &    0.24    &    -        &  40         & 4 \\
                 &    0.4     &   18.6      &  27         & 1 \\
                 &    0.4     &   19        &  $35\pm2$   & 2 \\
                 &    0.408   &    -        &  35         & 4 \\
                 &    0.41    &    -        &  40         & 5 \\
                 &    0.6     &   17.4      &  21         & 1 \\
                 &    0.61    &    -        &  55         & 4 \\
                 &    0.638   &   16        &  $37\pm2$   & 6 \\
                 &    0.92    &   16.4      &  24         & 1 \\
                 &    1.4     &   16.1      &  21         & 1 \\
                 &    1.5     &   15.1      &  $37.5\pm0.1$ & 13 \\
                 &    1.6     &   15.5      &  $18$         & 1 \\
    \end{tabular}
  \end{center}
\end{table}
%------------------------------------------------------------------------
\begin{table}
  \scriptsize
  \begin{center}
 %   \caption{table 1 continue---}
    \scriptsize
    \begin{tabular}{cllll}

                 &    1.612   &   16        &  $31\pm2$     & 7 \\
                 &    1.665   &    -        &  $30$         & 5 \\
                 &    1.72    &    -        &  $39.7\pm3$   & 8 \\
                 &    2.65    &    -        &  $21.3\pm5$   & 8 \\
                 &    4.85    &$14.2\pm0.4$ &  $9.2\pm1.6$  & 12 \\
                 &    10.55   &    -        &  $\leq5$      & 14 \\
      \hline
      J2022+5154 &    0.23    &    24.8     &  49         & 1  \\
                 &    0.392   &    -        &  50         & 5 \\
                 &    0.4     &    20.5     &  65         & 1  \\
                 &    0.6     &    19.7     &  45         & 1  \\
                 &    0.92    &    17.8     &  45         & 1  \\
                 &    1.4     &    16.9     &  40         & 1  \\
                 &    1.41    &    -        &  $28.6\pm0.1$ & 12 \\
                 &    1.6     &    17.4     &  32         & 1  \\
                 &    1.665   &    -        &  34         & 5 \\
                 &    1.71    &    -        &  $38.8\pm0.3$ & 12 \\
                 &    1.72    &    -        &  $39.2\pm6$   & 8 \\
                 &    2.65    &    -        &  $33\pm3$     & 8 \\
                 &    4.85    &    -        &  $22.7\pm0.3$ & 12 \\
                 &    8.75    &    -        &  $25.7\pm4$   & 8 \\
                 &    10.45   & $12.7\pm2.6$&  $29.6\pm6.3$ & 12 \\
                 &    10.5    &    -        &  $25\pm1$     & 14 \\
                 &    14.8    &    -        &  $15.9\pm6$   & 8 \\
      \hline
    \end{tabular}
  \end{center}
  Notes. $^\dagger$ The widths of PSR J0452-1759 are exceptionally
  measured at $50\%$ of the peak intensities. References for previous
  polarization observations: $^1$\citet{gl98}, $^2$\citet{hma+77},
  $^3$\citet{jkw06}, $^4$\citet{lsg71}, $^5$\citet{man71},
  $^6$\citet{mhm+78}, $^7$\citet{mhm80}, $^8$\citet{mgs+81},
  $^9$\citet{rb81}, $^{10}$\citet{vah+97}, $^{11}$\citet{vkk98},
  $^{12}$\citet{vx97}, $^{13}$\citet{wj08}, $^{14}$\citet{xsg+95}.
\end{table}

\label{lastpage}

\end{document}